\documentclass[aps, prl, twocolumn, showpacs, floatfix, superscriptaddress]{revtex4}
\usepackage{amsmath, amsfonts, amssymb}
\usepackage{bm}
\usepackage{graphicx}

\begin{document}

\newcommand{\odiff}[2]{\frac{\di #1}{\di #2}}
\newcommand{\pdiff}[2]{\frac{\partial #1}{\partial #2}}
\newcommand{\di}{\mathrm{d}}
\newcommand{\ii}{i}
\renewcommand{\vec}[1]{\bm{#1}}
\newcommand{\ket}[1]{|#1\rangle}
\newcommand{\bra}[1]{\langle#1|}
\newcommand{\pd}[2]{\langle#1|#2\rangle}
\newcommand{\tpd}[3]{\langle#1|#2|#3\rangle}

\title{Splitting of Majorana modes due to intervortex tunneling in a $p_x+i p_y$ superconductor}
\author{Meng Cheng}
\affiliation{Condensed Matter Theory Center and Joint Quantum Institute,
Department of Physics, University of Maryland, College Park, MD 20742}
\author{Roman M. Lutchyn}
\affiliation{Condensed Matter Theory Center and Joint Quantum Institute,
Department of Physics, University of Maryland, College Park, MD 20742}
\author{Victor Galitski}
\affiliation{Condensed Matter Theory Center and Joint Quantum Institute,
Department of Physics, University of Maryland, College Park, MD 20742}
\author{S. Das Sarma}
\affiliation{Condensed Matter Theory Center and Joint Quantum Institute,
Department of Physics, University of Maryland, College Park, MD 20742}

\date{
\today}

\begin{abstract}
We consider a two-dimensional $(p_x+i p_y)$-superconductor in the presence of multiple vortices, which support zero-energy Majorana fermion states in their cores. Intervortex tunnelings of the Majorana fermions lift the topological state degeneracy. Using the Bogoliubov-de Gennes equation, we calculate  splitting of the zero-energy modes due to these tunneling events. We also discuss superconducting fluctuations and, in particular, their effect on the energy splitting.
\end{abstract}


\maketitle

 Exotic excitations obeying non-Abelian statistics play a key role in topological quantum computation (TQC) and can emerge in a
 variety of condensed matter systems~\cite{dassarma_prl'05, DasSarma_PRB'06,  nayak_RevModPhys'08}, including chiral p-wave superfluids and
 superconductors~\cite{volovik_JETP'99, Read_prb'00}. Such non-Abelian quasiparticles are believed to be realized in the A-phase of superfluid
 $^3$He~\cite{Kopnin_PRB'91}, the oxide superconductor $\rm Sr_2RuO_4$~\cite{ivanov_prl'01}, p-wave superfluids in cold atom settings~\cite{Gurarie_annals} as well as at the interfaces of s-wave superconductor and topological insulator~\cite{fu_prl'08}.
 Certain vortex excitations in these systems support zero-energy Majorana fermions residing inside their cores, which lead to topological degeneracy
 of the ground states and non-Abelian statistics when multiple spatially separated vortices are present~\cite{Read_prb'00, ivanov_prl'01}.

In a spin-triplet $(p_x+ip_y)$-superconductor, the zero-energy states appear in half-quantum vortices,  where the complex phase of the condensate wavefunction  winds in one of the spin-sectors, \emph{i.e.} $\Phi(r,\varphi)\!=\!\Delta(r)(p_x+i p_y) [\ket{\!\downarrow\downarrow}+e^{i\varphi}\ket{\uparrow \uparrow}]$.  This effectively is equivalent to a full-quantum vortex in a spinless $(p_x+ip_y)$-superfluid/superconductor, which too may host a zero-energy state~\cite{ivanov_prl'01}. The existence of  such states has been well established through several approaches, including quasi-classical analysis~\cite{volovik_JETP'99}, explicit solutions of Bogoliubov-de Gennes(BdG) equation~\cite{Kopnin_PRB'91, stone_prb'06, Gurarie_prb'07} and an index theorem~\cite{Tewari_prl'07}. These zero energy states can be occupied by Majorana  fermions and lead to a degeneracy of the many-body ground state, which hinges on the exact degeneracy of the single Majorana modes in different vortex cores. However, in  the presence of multiple vortices, tunneling  of the Majorana modes becomes possible and these tunneling events are expected to lift the ground-state degeneracy to some degree~\cite{nayak_RevModPhys'08}.

For the purposes of topological quantum computation, it is crucial to understand the stability of Majorana modes against different perturbations such as fluctuation effects and the intervortex tunneling processes. In this Letter we concentrate on the latter tunneling effects in a two-dimensional spinless $(p_x + ip_y)$-superconductor and calculate the energy splitting of the Majorana modes, including its dependence on the distance between vortex cores. We find that in addition to an exponential suppression $\exp(-R/\xi)$, where $R$ and $\xi$ are the intervortex distance and the superconducting coherence length, respectively,  the amplitude of the tunneling rate oscillates on the scale of the Fermi wavelength, see Eq.~(\ref{eq:splitting}). These oscillations have important consequences for TQC which we discuss below. We also study thermal motion of a vortex which leads to smearing out the fast-oscillating term. We calculate the typical value of the degeneracy splitting (defined by its root-mean-square value) which determines the ``decoherence'' time for TQC.    

{\it Theoretical model.}  Our starting point is the mean-field BCS Hamiltonian of spinless $p_x+i p_y$ superconductor~\cite{Stern_prb'04} 
\begin{multline}
\mathcal{H}_{\mathrm{BCS}}=\int \di^2\vec{r}\, \hat{\psi}^\dag(\vec{r})\Big(-\frac{\hbar^2\nabla^2}{2m}-\mu\Big)\hat{\psi}(\vec{r})\\
+\frac{i \hbar }{p_F}\int \di^2\vec{r}\,\Delta(\vec{r})\pdiff{\hat{\psi}^\dag(\vec{r})}{\overline{z}}\hat{\psi}^\dag(\vec{r})+\mathrm{h.c.},
\label{eq:bcs}
\end{multline}
which enforces the $p_x+i p_y$  symmetry for the superconducting gap: $\Delta(\vec{p})=\frac{\Delta_0}{p_F}(p_x+i p_y)$. Here $z=x+i y$ is complex coordinate in the 2D plane. The BdG equation follows from diagonalizing Hamiltonian \eqref{eq:bcs} via Bogoliubov transformation $\hat{\psi}(\vec{r})=\sum_n \big[\hat{\gamma}_nu_n(\vec{r})+\hat{\gamma}_n^\dag v_n^\ast(\vec{r})\big]$, and has the form (henceforth we use $\hbar=k_B=1$)
\begin{align}
&\mathcal{H}_{\mathrm{BdG}}
\begin{pmatrix}
u_n(\bm r)\\
v_n(\bm r)
\end{pmatrix}=E_n\begin{pmatrix}
u_n(\bm r)\\
v_n(\bm r)
\end{pmatrix}, \label{eq:bdg}\\
\mathcal{H}_{\mathrm{BdG}}&\!=\! \begin{pmatrix}
{\displaystyle -\frac{\nabla^2}{2m}\!-\!\mu} & {\displaystyle\frac{i}{p_F}\Big\{\Delta(\vec{r}), \pdiff{}{\overline{z}}\Big\}}\\
{\displaystyle \frac{i}{p_F}\Big\{\Delta^\ast(\vec{r}),\pdiff{}{z}\Big\}} & {\displaystyle\frac{\nabla^2}{2m}+\mu}
\end{pmatrix}.
\label{eq:Hbdg}
\end{align}
The Hamiltonian $\mathcal{H}_{\mathrm{BdG}}$ is invariant under transformation, $\sigma_1 \mathcal{H}_{\mathrm{BdG}}\sigma_1=-\mathcal{H}_{\mathrm{BdG}}^\ast$ , which relates solutions with positive and negative energies~\cite{Gurarie_prb'07}. This symmetry implies that if $\Psi=(u_n,v_n)^T$ is a solution of Eq.~(\ref{eq:bdg}) with eigenvalue $E_n$, then $\sigma_1\Psi^\ast=(v_n^\ast,u_n^\ast)^T$ must be a solution with the eigenvalue $(-E_n)$. Thus, for a zero-energy state, we have the constraint  $u=v^\ast$. In the presence of a vortex this ensures the existence of a stable, symmetry-protected zero-energy state. Similar to s-wave superconductors~\cite{Caroli_PL'64}, the $hc/2e$ vortex can be modeled as $\Delta(\vec{r})=f(r)e^{\ii \varphi}$, where $\varphi$ is the polar angle and $f(r)$ is the superconducting order-parameter profile of a vortex, $f(r)=\Delta_0 \tanh\left(\frac{r}{\xi}\right)$ with $\Delta_0$ being the mean-field value of the superconducting order parameter. By directly solving BdG equation \eqref{eq:bdg}, one finds the energy spectrum for the bound states in the vortex core as $E_n= - \omega_0 n$, where $n$ is an integer and
 $\omega_0\sim\Delta_0^2/\varepsilon_F$ with $\varepsilon_F$ being the Fermi energy~\cite{Kopnin_PRB'91}. The eigenstate corresponding to $n=0$ is given by~\cite{tewari_prl'2007, Gurarie_prb'07}
\begin{equation}
\!\Psi(\vec{r})\!=\!\!\sqrt{\frac{p_F}{2\pi\xi}}J_1(p_Fr)\!\exp\!\left[\!\ii \left(\varphi-\frac{\pi}{4}\right)\sigma_3\!-\!\frac{1}{v_F}\!\!\int_0^r\! \di r'f(r')\!\right]\!,
\label{onevortex}
\end{equation}
where $v_F=p_F/m$ is the Fermi velocity, and $J_1
(r)$ is the Bessel function. The constant phases of $u$ and $v$ are chosen to satisfy the requirement that $u^\ast=v$. Eq.~\eqref{onevortex} was obtained assuming $\Delta_0\ll \varepsilon_F$ which is typical for weak-coupling superconductors. Using the zero-mode solution $(u,v)^T$ with $u=v^\ast$, we can construct the Majorana quasiparticle operator $\hat{\gamma}=\hat{\gamma}^\dagger = \int \di^2\vec{r}\,\big[\hat{\psi}(\vec{r})u^\ast(\vec{r})+\hat{\psi}^\dag(\vec{r})v^\ast(\vec{r})\big]$.

Let us now consider the situation with $2N$ vortices pinned at positions $\vec{R}_i$. If we ignore the fluctuation effects and the tunneling events, the superconducting order parameter can be represented as $\Delta(\vec{r})=\prod_{i=1}^{2N}f(\vec{r-R_i})\exp\big[\ii\sum_i\varphi_i(\vec{r})\big]$, where $\varphi_i(\vec{r})=\mathrm{arg}(\vec{r}-\vec{R}_i)$. Near the $k$-th vortex core, the phase of the order parameter can be approximated by $\varphi_k(\vec{r})+\Omega_k$ with $\Omega_k=\sum_{i\neq k}\varphi_i(\vec{R}_k)$. Thus, one can generalize the zero-energy solution obtained for a single vortex to the situation at hand~\cite{Stern_prb'04}:
\begin{align}
\Psi_i(\vec{r})&=\sqrt{\frac{p_F}{2\pi\xi}}J_1(p_Fr_i)\exp\left[-\frac{1}{v_F}\int_0^{r_i} \di r'f(r')\right]\nonumber\\
& \times \exp\left[ \ii \left(\varphi_i+\frac{\Omega_i}{2}-\frac{\pi}{4}\right)\sigma_3\right].
\label{zeromode2}
\end{align}
The $2N$ Majorana fermions residing in the vortex cores can be combined in pairs to create $N$ Dirac fermions, $\hat{c}=\frac{1}{\sqrt{2}}(\hat{\gamma}_i+\ii\hat{\gamma}_j),
\hat{c}^\dag= \frac{1}{\sqrt{2}}(\hat{\gamma}_i-\ii\hat{\gamma}_j)$. This allows one to enumerate all degenerate ground states,
which can be occupied by $N$ fermions~\cite{nayak_RevModPhys'08}.

\begin{figure}
\begin{center}
\includegraphics[width=0.8\linewidth]{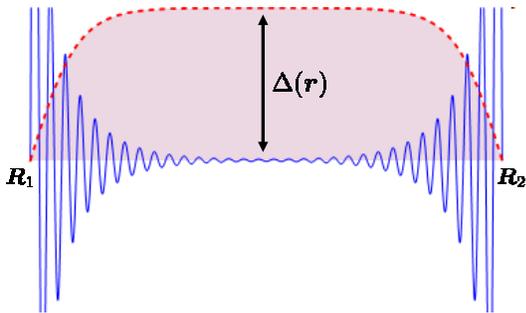}
\end{center}
\caption{(color online). Schematic plot of a two-vortex configuration. The shaded region corresponds the order parameter profile $f(\bm r)$ with vortices located at $\bm R_1$ and $\bm R_2$. The solid (blue) line represents the real part of the wavefunction $u_1(\bm r)+i u_2(\bm r)$.}
\label{Fig:majorana}
\end{figure}

To understand the nature of the states created by the pairs of Majorana fermions, we  focus here on the case of two vortices residing at the positions $\vec{R}_1$ and $\vec{R}_2$. Formally, a vortex potential in Eq.~(\ref{eq:bdg}) is similar to a quantum well, in which the core excitations including the Majorana modes may reside. The two-vortex case is, therefore, similar to a double-well problem in which the tunneling events are expected to lead to a splitting of the originally degenerate energy levels~\cite{Landau_book3}. As a result, the degeneracy between Dirac fermion states $\hat{c}^\dag \hat{c} \ket{1}=\ket{1}$ and $\hat{c}^\dag \hat{c} \ket{0}=0$ is lifted by the tunneling between the two vortices. Here the Dirac fermion $\hat{c}=(\hat{\gamma}_1+i\hat{\gamma}_2)/\sqrt{2}$ is constructed out of two Majorana fermions $\gamma_1$ and $\gamma_2$. The energy difference between these two states is the main quantity of interest in this work. Similarly to a double-well problem, we first find the single-vortex solutions at $\bm R_1$ and $\bm R_2$: $\Psi_1=(u_1,v_1)^T$ and $\Psi_2=(u_2,v_2)^T$, and then construct two-vortex wavefunctions 
$\Psi_\pm=\left(\Psi_1 \pm e^{i\alpha}\Psi_2\right)/\sqrt{2}$, which correspond to the energies $E_{+}$ and $E_{-}$, respectively. The energies $E_{+}$ and $E_{-}$ are related by $E_{+}=-E_{-}$. The particle-hole symmetry of the BdG equations requires that  
$\sigma_1 \Psi^*_+=\Psi_-$. Combining this constraint with the properties of the zero energy solutions $u=v^*$, we find that $\alpha=\pi/2$. Consequently, the Dirac fermion operators $\hat{c}$ and $\hat{c}^\dag$ can be identified as annihilation and creation operators of the single-particle state $\Psi_+$, \emph{i.e}
\[
\hat{c}=\frac{\hat{\gamma}_1+\ii\hat{\gamma}_2}{\sqrt{2}}=\int \di^2\vec{r}\,\left[\hat{\psi}\frac{u_1^\ast+\ii u_2^\ast}{\sqrt{2}}+\hat{\psi}^\dag\frac{v_1^\ast+\ii v_2^\ast}{\sqrt{2}}\right].
\]
Thus, $E_+-E_-=2E_+$ is simply the energy splitting between the occupied state $\ket{1}$ and unoccupied state $\ket{0}$~\cite{stone_prb'06}. The energy of the state $\Psi_+$, for example, can be calculated as the appropriate overlap integral between two zero-energy states~\cite{Landau_book3}. When the distance between two vortices is much larger than superconducting coherence length $R=|\bm R_2-\bm R_1| \gg \xi=v_F/\Delta_0$, the wavefunction $\Psi_2$ is exponentially small in region close to the vortex at $\bm R_1$. Then, the splitting energy $E_+$ is given by
\begin{align}\label{eq:overlap}
E_+=\frac{\int_\Sigma \di^2\vec{r}\,\Psi_1^\dag\mathcal{H}_{\mathrm{BdG}}\Psi_+-\int_\Sigma \di^2\vec{r}\,\Psi_+^\dag\mathcal{H}_{\mathrm{BdG}}\Psi_1}{\int_\Sigma\di^2\vec{r}\,\Psi_1^\dag\Psi_+}.
\end{align}
Here $\Sigma$ is the half plane $x \in (0,\infty), y \in (-\infty,\infty)$ containing one of the vortex at $\bm R_1=(R/2,0)$. The other vortex is located at $\bm R_2=(-R/2,0)$. The integral in Eq.~(\ref{eq:overlap}) can be calculated using the explicit form of the solution for $\Psi_1$ and $\Psi_+$.
We first transform integral in Eq.~(\ref{eq:overlap}) over half-plane into a line integral over the boundary of $\Sigma$ at $x=0$: 
\begin{equation}
E_+ \approx - \frac{\sqrt{2}\Delta_0 a}{\pi^{2}}\!\int_{-\infty}^{\infty}\!\!dy\!\, \frac{\cos \!\left(\!2\lambda \sqrt{a^2\!+\!y^2}\right)}{a^2\!+\!y^2}\,\!\exp\!\left(\!-2\sqrt{a^2\!+\!y^2}\right)\nonumber\!,
\end{equation}
where $a=R/2\xi$ and $\lambda=p_F\xi$. Upon evaluating the integral above, we find the splitting energy to be
\begin{equation}
E_+ \approx  - \frac{2\Delta_0}{\pi^{3/2}}\frac{\cos\big(p_FR+\frac{\pi}{4}\big)}{\sqrt{p_FR}}\,\exp\left(-\frac{R}{\xi}\right).
\label{eq:splitting}
\end{equation}
Here we neglect the corrections to the prefactor of order $(p_F\xi)^{-1}\ll 1$. As one can see from Eq.~(\ref{eq:splitting}), in addition to the expected exponential decay, the splitting energy oscillates rapidly on the Fermi wavelength. These oscillations originate from the quantum interference between two zero-energy eigenstates located at $\bm R_1$ and $\bm R_2$, see Fig~\ref{Fig:majorana}. The analytical result for the degeneracy splitting \eqref{eq:splitting} is in qualitative agreement with recent numerical studies of the topological degeneracy in the quantum Hall state at Landau level filling fraction $\nu=5/2$~\cite{Tserkovnyak_PRL'03} as well as in Kitaev's honeycomb lattice model~\cite{Lahtinen_annals'08}. We note in passing here that the honeycomb lattice model can be mapped to a p-wave superconductor. However, the elementary topological excitations in the Kitaev's model (vortices living on a plaquette) are different from half-quantum vortices in the chiral p-wave superconductors although both have non-Abelian Ising anyons.

In the discussion above we assumed zero temperature limit. At finite temperature, the error rate also comes from thermal fluctuations~\cite{dassarma_prl'05}. To take advantage of the topological quantum computation, the temperature should be smaller than the excitation gap, which in p-wave superconductors is given by the level spacing of bound states in the core, $\omega_0 \sim \Delta_0^2/E_F$. At $T\ll \omega_0$, thermal population of the excited bound states in the vortex core is exponentially small $\propto \exp\left(-\omega_0/T\right)$, and, thus, the error rate is suppressed. In addition to this well known argument~\cite{dassarma_prl'05}, finite temperature leads to fluctuation effects discussed in the remainder of the paper.
\newcommand{\rv}{{\mathrm{v}}}

{\it Fluctuation effects.} The BdG equation (\ref{eq:bdg}) follows from the mean-field Hamiltonian (\ref{eq:bcs}) and assumes that the superconducting order parameter  is a fixed field with no internal dynamics. This concerns both the uniform  order-parameter background $\Delta_0$ and the vortex field, which in Eq.~(\ref{eq:bdg}) is assumed to be an externally imposed defect pinned in a certain location. However, the mean-field Hamiltonian is an approximation and there are corrections to it due to classical and quantum fluctuations of the order parameter which can be separated into amplitude and phase fluctuations. The amplitude fluctuations manifest themselves most strongly near the transition and, even though they do exist in the superconducting phase as well, they are well-gapped and their effects are not as dramatic there. The bulk phase fluctuations are gapless in two dimensions both in a neutral system and in a charged superfluid, and, thus, can propagate over large distances. However, if the film thickness is finite, the phase fluctuations do not propagate beyond a certain magnetic screening length and hence become local~\cite{mpa_prl'90}.   If a vortex is present, the local phase fluctuations effectively restore the vortex  dynamics and, in particular, give rise to spatial motion of the vortex position as opposed to a fixed static configuration assumed in Eq.~ (\ref{eq:bdg}). This motion as well as any other {\em local} perturbations are unlikely to destroy a single topological Majorana mode, but the vortex motion  certainly has a strong  effect on the Majorana mode splitting, as can be seen from Eq.~(\ref{eq:splitting}). Indeed, the splitting contains the fast-oscillating function $\cos\big(p_FR+\frac{\pi}{4}\big)$ and as such is extremely sensitive to the vortex positions. The Fermi wavelength is by far the smallest lengthscale in the problem and motion of the vortex effectively ``smears out'' the fast-oscillating term, introducing a new lengthscale in the problem associated with vortex dynamics.

 In the superfluid state at zero temperature, vortex dynamics is quantum, while at higher temperatures the fluctuations of the vortex position becomes effectively classical and have a timescale associated with it. To illustrate the main effect of fluctuations, we adopt here a simple  phenomenological model of a single vortex moving a pinning potential, $V(\vec{R}_\rv)$~\cite{Niu_prl'94, Bartosch_PRB'06}
\begin{equation}
\mathcal{H}_{\mathrm{vortex}}=\frac{1}{2m_\rv}(\hat{\vec{p}}_\rv-\vec{\mathcal{A}})^2+V(\hat{\vec{R}}_\rv).
\label{eq:vortex_hamiltonian}
\end{equation}
Here $\hat{\vec{p}}_\rv$ and $\hat{\vec{R}}_\rv$ are the canonical momentum and coordinate of the vortex core, respectively,  $m_\rv$  is the vortex effective mass, determined by the virtual transitions among the quasiparticle states caused by the vortex motion which is treated as a phenomenological parameter below. Eq.~(\ref{eq:vortex_hamiltonian}) also  includes effective gauge field $\vec{\mathcal{A}}$ due to the Magnus force, which modifies the vortex motion, however we ignore it in actual calculations of fluctuation vortex dynamics. Assuming the displacement of the vortex position from equilibrium is small, one can approximate the potential $V(\vec{R}_\rv)$ by a harmonic ``pinning trap'': $V(\vec{R}_\rv)\approx\frac{1}{2}m_\rv\omega^2_\rv(\vec{R}_\rv-\vec{R}_{\rv 0})^2$, where $\vec{R}_{\rv 0}$ is the equilibrium position.  The total Hamiltonian, which accounts for the coupling between vortex motion and the quasiparticles residing in the vortex core reads~\cite{Bartosch_PRB'06}
\begin{equation}
\mathcal{H}=\mathcal{H}_{\mathrm{vortex}}+\mathcal{H}_{\mathrm{BCS}}(\{\hat{\vec{R}_\rv}\}),
\label{hamiltonian_full}
\end{equation}
where $\mathcal{H}_{\mathrm{BCS}}$ is the BCS Hamiltonian \eqref{eq:bcs}. The order parameter $\Delta(\vec{r})$ in Eq. \eqref{eq:bcs} now depends on the vortex positions $\{\hat{\vec{R}}_\rv\}$. Below, we consider the temperature regime $\omega_\rv\ll T\ll\omega_0$, where the fluctuations of the vortex position are dominated by classical fluctuations. In this parameter regime the motion of the vortex is slow compared to the time scale of quasiparticle dynamics inside the core, $ \omega_0^{-1}$. Therefore,  the solution for the zero-energy eigenstates remains valid, and the calculation of the splitting energy \eqref{eq:splitting} goes through as before. However, the energy splitting, $E_+$, itself becomes a random variable since it depends on the intervortex separation, which fluctuates.  In general, we are interested in the full probability distribution function of the eigenvalue splitting given by $P[E_+] = \left\langle \delta\left[ E_+ - E_+(\vec{R}_1,\vec{R}_2) \right] \right\rangle$, where the average of a function $f(\vec{R}_1,\vec{R}_2)$ is defined as the integral over the vortex positions $\langle f \rangle =\int\di^2\vec{R}_1\di^2\vec{R}_2 f(\vec{R}_1,\vec{R}_2)\rho_1(\vec{R}_1)\rho_2(\vec{R}_2)$ weighted with the diagonal element of the density matrix of a harmonic oscillator:
$\rho(\vec{R})=
\exp\! \left( \! -\vec{R}^2/l^2\right)/\pi l^2$,
where $l=1/\sqrt{m_\rv \omega_\rv \tanh (\beta \omega_\rv / 2)}$ represents typical deviation of the vortex position from the equilibrium. Below we calculate explicitly the average Majorana mode splitting, $\langle E_+\rangle$,  and its root-mean-square value, $\sqrt{\langle E_+^2\rangle}$.
In the limit $\omega_\rv \ll T$, $l$ is asymptotically given by $l\sim \sqrt{2 T/ m_\rv \omega_\rv^2}$. Using the value for the vortex mass of Refs.~\cite{Kopnin_PRB'91, Bartosch_PRB'06}, we estimate that in the physically relevant regime $p_F^{-1} \ll l \ll \xi \ll R$ the expression for $\langle E_+\rangle $ becomes
\begin{equation}\label{eq:sp_aver2}
\langle E_+\rangle\! \approx \! - \! \frac{2\Delta_0}{\pi^{\frac 3 2}}\frac{\!\cos\!\big[p_F R\!+\!\frac{\pi}{4}\big]}{\sqrt{p_FR}}
\,\!\exp\!\left(-\frac{R}{\xi} -\frac{p_F^2l^2}{2}\right)\!.
\end{equation}
Compared with Eq.\eqref{eq:splitting}, the averaged splitting rate becomes exponentially smaller than its ``static'' value (\ref{eq:splitting}) due to the vortex position fluctuations. This smallness originates from the cosine function in Eq.~(\ref{eq:splitting}), which oscillates in sign and whose average value is therefore small. One may, therefore, conclude naively that the relevant suppression parameter is large since $p_Fl \gg 1$. This effect is similar to the ``nominal'' suppression of the Friedel oscillations and RKKY interactions in disordered metals~\cite{Spivak} and superconductors~\cite{GaL}. In the case of RKKY interaction it is well-known that the latter \emph{ensemble-averaged} result is unphysical and the typical value of Friedel interaction terms is not small~\cite{Spivak}. In our case the situation is similar and this is illustrated by the root-mean-square value
\begin{equation}\label{eq:rmsq}
\sqrt{\langle E_+^2\rangle} \approx \frac{\sqrt 2 \Delta_0}{\pi^{\frac 3 2}}\frac{\exp\left(-R/\xi\right)}{\sqrt{p_FR}},
\end{equation}
which much exceeds the average (\ref{eq:sp_aver2}), $\langle E_+\rangle/\sqrt{\langle E_+^2\rangle} \sim \exp\left(-\frac{1}{2}p_F^2l^2\right)$. This result
for $\sqrt{\langle E_+^2\rangle}$ represents a {\em typical} value of the splitting in the parameter regime of $R \gg \xi$ which sets the upper limit for the duration of the braiding operations in TQC. 
   
The dependence of the degeneracy splitting on physical parameters is important for TQC since it sheds light on both the adiabaticity of braiding operations and decoherence. In particular, the magnitude of the splitting energy sets the upper limit for the time of braiding operations which have to be performed adiabatically with respect to the excitation gap in the system $\omega_0$. On the other hand, because of the lifting of the topological state degeneracy the statistical phase gets smeared out at times $t \gtrsim 1/\sqrt{\langle E_+^2\rangle}$. Thus, the non-Abelian statistics of the Majorana excitations can be resolved for a wide range of time scales $\omega_0^{-1} \ll t \ll 1/\sqrt{\langle E_+^2\rangle}$ as long as the vortices well-separted so that the splitting energy remains exponentially small.  Furthermore, the sign of the splitting energy can be either positive or negative, energetically favoring unoccupied or occupied state by the Dirac fermion. Because of the oscillations of the splitting energy on the atomic length scale, the initialization of the qubit in the desired quantum state as well as the read-out based on bringing two vortices together would become difficult. 

 We conclude by pointing out that our work has important consequences for topological quantum computation using non-Abelian anyonic degeneracy as proposed in several related architectures recently~\cite{dassarma_prl'05, DasSarma_PRB'06, fu_prl'08, tewari_prl'2007}. The energy splitting we calculate will suppress topological immunity since it destroys the exact degeneracy of the quasiparticle subspace. This may adversely affect the fault tolerant properties of the topological quantum computation. Our work is the first analytical theory of quantum decoherence in topological quantum computation demonstrating the limitation of the topological protection due to quasiparticle tunneling. 

This work was supported by DARPA-QuEST.


\end{document}